\documentclass[11pt,preprint,graphicx]{aastex}

\usepackage[fleqn]{amsmath} 
\usepackage{amsfonts} 
\usepackage{amstext}
\usepackage{amssymb} 
\usepackage[colorlinks=true, citecolor=blue, linkcolor=blue, breaklinks=true]{hyperref}
\usepackage{natbib}
\usepackage{graphicx}

\def\etal{\it et al. \rm }

\begin{document}

\title{Using The Baryonic Tully-Fisher Relation to Measure $H_o$}

\author{James Schombert$^{1}$,Stacy McGaugh$^{2}$ and Federico Lelli$^{3}$}
\affil{$^{1}$Institute for Fundamental Science, University of Oregon, Eugene, OR 97403}
\affil{$^{2}$Department of Astronomy, Case Western Reserve University, Cleveland, OH 44106}
\affil{$^{3}$School of Physics and Astronomy, Cardiff University, Queens Buildings, The
Parade, Cardiff, CF24 3AA, UK}

\begin{abstract}

\noindent We explore the use of the baryonic Tully-Fisher relation (bTFR) as a new
distance indicator.  Advances in near-IR imaging and stellar population models, plus
precise rotation curves, have reduced the scatter in the bTFR such that distance is
the dominant source of uncertainty.  Using 50 galaxies with accurate distances from
Cepheids or tip magnitude of the red giant branch, we calibrate the bTFR on a scale
independent of $H_o$.  We then apply this calibrated bTFR to 95 independent galaxies
from the SPARC sample, using CosmicFlows-3 velocities, to deduce the local value of
$H_o$.  We find $H_o = 75.1 \pm 2.3$ (stat) $\pm 1.5$ (sys) km s$^{-1}$ Mpc$^{-1}$.

\end{abstract}

\section{Introduction}

The classic Tully-Fisher (TF) relation links the rotation velocity of a disk galaxy
to its stellar mass and/or luminosity in a given photometric band. Because the observed
rotation velocities do not depend on galaxy distance $D$, while stellar luminosities
depend on $D^2$, the TF relation is a key distance indicator that has played a
crucial and historical role in constraining the value of $H_o$ (see Tully \& Fisher
1977, Sakai \etal 2000). The classic TF relation, however, breaks down at stellar
masses below approximately 10$^9$ M$_\odot$, when dwarf galaxies in groups and in the
field environment become progressively more gas rich. By replacing the stellar mass
with the total baryonic mass (stars plus gas, $M_b$), one recovers a single linear
relation: the so-called baryonic Tully-Fisher relation (bTFR, Freeman 1999; McGaugh
\etal 2000; Verheijen 2001, {Zaritsky \etal 2014}).

The bTFR samples deeper into the galaxy mass function, as low-mass dwarfs typically
have high gas fractions and the neutral gas can constitute 80 to 90\% of the total
baryonic mass (Bradford \etal 2015).  This results in a large amount of scatter in
the classic TF relation with a corresponding loss in accuracy as a distance
indicator, which is eliminated by including the gas mass. Currently, the bTFR extends
over two decades in velocity and six decades in $M_b$ (McGaugh 2012, Iorio \etal
2017).  Moreover, it displays a surprisingly small scatter, considering the number of
possible competing astrophysical processes that produce this relation (Lelli, McGaugh
\& Schombert 2016a). 

The bTFR combines rotational velocity ($V$) and baryon mass ($M_b$) in the form of
$M_b = AV^x$  that, unlike most astrophysical correlations, does not show any room
for a third parameter such as characteristic radius or surface brightness (Lelli
\etal 2019).  In a log-log bTFR plot, $x$ becomes the slope of the relationship, and log
$A$ becomes the zeropoint.  As $V$ is distance independent, errors in distance only
reflect into $M_b$ (going as $D^2$).  While the scatter in the bTFR can be used to
constrain galaxy formation models in a $\Lambda$CDM cosmology (Dutton 2012), it also
presents a unique opportunity to test the consistency of the distance scale
zero-point (i.e., $H_o$) with redshift-independent calibrators of the bTFR.

A new local test of $H_o$ has become critical, for over the past decade there have
been growing discrepancies in the determination of $H_o$ by different methods.  On
one hand, the Cepheid (C) calibration to type-Ia supernovae (SN) yields
$H_o=73.2\pm1.7$ km s$^{-1}$ Mpc$^{-1}$ (Riess \etal 2016).  Using the tip of the red
giant branch (TRGB) method to calibrate the SN distance scale yields a slightly lower
$H_o=69.8\pm0.8$ km s$^{-1}$ Mpc$^{-1}$ (Freedman \etal 2019).  However, fitting the
angular power spectrum of cosmic microwave background (CMB) fluctuations in the
Planck data with $\Lambda$CDM models produces a value of $H_o$ of $67.4\pm0.5$
(Planck Collaboration 2016). In addition, an $H_o$ value of $67.3\pm1.1$ was found
from the SDSS-III Baryon Oscillation Spectroscopic Survey (BOSS) measurements of the
baryon acoustic oscillations (Aubourg \etal 2015). Both these measurements, which
hinge on a $\Lambda$CDM-driven interpretation of the CMB, represent a 6$\sigma$
difference from the Cepheid-calibrated SN distance scale (Verde, Treu \& Riess 2019).

In this paper, we present a redshift-independent calibration of the bTFR using
galaxies with Cepheid and TRGB distances (see Sorce \etal 2013) from a subset of the
$Spitzer$ Photometry and Accurate Rotation Curves (SPARC) sample and 
of Ponomareva \etal (2018).  This baseline bTFR is then compared
to the larger SPARC data set with new CosmicFlows-3 velocities (Tully \etal 2019) to
examine the variations in the bTFR normalization for different $H_o$ values.

\section{Data}

\subsection{The SPARC galaxy database}

The SPARC data set consists of HI rotation curves (RC) accumulated over the last
three decades of radio interferometry combined with deep near-IR photometry from the
$Spitzer$ 3.6$\mu$m IRAC camera (Lelli, McGaugh \& Schombert 2016b).  This provides
the community an important combination of extended HI rotation curves (mapping the
galaxy gravitational potential out to large radii) plus near-IR surface photometry to
map the stellar component (see also Zaritsky \etal 2014).  In addition, the HI
observations also provide the HI gas mass that, when corrected for small amounts of He
and heavier elements, becomes the total gas mass of a galaxy.

The SPARC sample spans a broad range in baryonic mass (10$^8$ to 10$^{11}$
$M_{\odot}$), surface brightness (3 to 1000 $L_{\odot}$ pc$^{-2}$) and rotation
velocity ($V_{f}$ from 20 to 300 km sec$^{-1}$).  The SPARC dataset also contains
every Hubble late-type producing a representative sample of disk galaxies
from dwarf irregulars to massive spirals with large bulges.  The details of the
sample are listed in Lelli, McGaugh \& Schombert (2016b) and the resulting science
outlined in McGaugh, Lelli \& Schombert (2016).  The SPARC sample and analysis with
respect to the bTFR are presented in Lelli, McGaugh \& Schombert (2016a).  The analysis
presented herein follows that paper with respect to error analysis plus small
additions and corrections to the data as outlined in Lelli \etal (2019).

For sample selection, we have isolated a subset of 125 galaxies that follow the
quality criterion outlined in Lelli \etal (2019) plus two additional galaxies with
good TRGB distances.  That study examined 153 objects in the original sample,
excluding 22 galaxies with low inclinations ($i < 30$) where geometric corrections
are uncertain, and another six galaxies with low-quality rotation curves that do not
appear to track the equilibrium gravitational potential of the system.  As discussed
in Lelli \etal (2019), the average circular velocity along the flat portion of the
rotation curve ($V_f$) results in the tightest correlation between rotation and total
baryonic mass (see their Figure 2).  This measure of rotation velocity is superior to
single-dish measures (such as $W_{P20}$ or $W_{M50}$) or velocities based on some
disk scale length (e.g., $V_{2R_e}$) or peak of the rotation curve.  {Only those
rotation curves that display a flat outer portion (neither rising or falling) were
included in the sample, and it is those values ($V_f$), that will form one axis of the
bTFR for our analysis.}

The other axis of the bTFR is total baryonic mass, the sum of all of the observed
components, stars and gas ($M_b = M_* + M_g$).  Of the two components to the baryonic
mass, the stellar component has the highest uncertainty as it is determined by
measuring a luminosity at a specific wavelength multiplied by the appropriate
mass-to-light ratio ($\Upsilon_*$) for that wavelength.  The mass-to-light ratio is
obtained from stellar population models (e.g., Bell \etal 2003, Portinari \etal 2004,
Meidt \etal 2014, Schombert, McGaugh \& Lelli 2019) considering the galaxy color to
account for the effects of star formation.  The physics of star formation, and later
stellar evolution, are such that the $\Upsilon_*$ deduced from optical luminosities are
highly sensitive to the galaxy's star formation rate (SFR) and produce uncertain
stellar masses.  Values in the near-IR are less sensitive to a galaxy's star
formation history and have the additional advantage of minimizing absorption by dust.
For this analysis, we use the $\Upsilon_*$ values obtained from Schombert, McGaugh \&
Lelli (2019) of 0.5 for disk regions and 0.7 for bulge regions at the IRAC channel 1
wavelength of 3.6$\mu$m.  We use fits to the $Spitzer$ 3.6 surface brightness
profiles to determine a galaxy's bulge-to-disk ratio ($B/D$) and apply the
appropriate $\Upsilon_*$ to the luminosities of those components.

The SPARC sample pays extra attention to gas-rich, low surface brightness (LSB)
galaxies that typically populate the low-mass end of the bTFR.  Unlike the luminous
TF (Tully \etal 2013), the galaxies on the low-mass end are dominated by a gas
component, and simply using the galaxy luminosity as a proxy for baryonic mass is
inaccurate.  Fortunately, the same HI observations that yield the rotation velocity,
$V_f$, also provide detailed information about the gas content.  In particular,
neutral atomic (HI) gas dominates the gas component in typical star-forming galaxies
and, therefore, the gas mass can be deduced directly from the physics of the
spin-flip transition of hydrogen times the cosmic hydrogen fraction plus minor
corrections for molecular hydrogen and heavier elements.  While the contribution of
ionized gas is considered in numerical simulations (Gnedin 2012), we found no
evidence of large amounts of diffuse H$\alpha$ emission or X-ray output in the
gas-rich galaxies of our sample (Schombert \etal 2011; Qu \& Bregman 2019) where such
a correction would dominate.

\subsection{Uncertainties}

Aside from distance errors, errors to the baryon mass have four components: (1)
errors in the 3.6 photometry, (2) errors in the HI fluxes, (3) uncertainty in
the conversion of near-IR luminosity to stellar mass ($\Upsilon_*$) and (4) uncertainty
in the conversion of atomic gas mass into total gas mass.  The first two are due to
errors in the observations and are outlined in Lelli, McGaugh \& Schombert (2016b).
These error estimates are substantiated by comparison to 2MASS $K$-band photometry
and other HI studies (Schombert \& McGaugh 2014).  They are typically of the order of
3\% for the 3.6 luminosities and 10\% for HI fluxes, which translate into a mean
error of 0.04 in log $M_b$.

The second set of errors are systematic to assumptions in the conversion of the
observed values to mass values.  For example, systematic errors in the modeling of
$\Upsilon_*$ are explored in Schombert, McGaugh \& Lelli (2019) where scenarios with
different assumptions on the star formation history or the stellar mass function produced different
relationships between color and $\Upsilon_*$.  However, there are only a limited
number of stellar population models that also reproduce the main-sequence diagrams
(the correlation of stellar mass versus star formation rate which indicate nearly
constant SF over a Hubble time for most of the SPARC sample) and the distribution of
colors from the UV to near-IR (see Schombert, McGaugh \& Lelli 2019).  Those models
are well approximated by using a singular value of 0.5 for $\Upsilon_*$ in the disk
regions and a value of 0.7 for bulges.

Through the use of Bayesian rotation-curve fits with dark matter halos, the plausible
galaxy-to-galaxy variations of $\Upsilon_*$ can be explored (Li \etal 2020).  From
those simulations, we find that the $\Upsilon_*$ at 3.6$\mu$m for the blue colors
typical of low-mass galaxies span a range from 0.45 to 0.60.  This is in agreement
with the possible range of $\Upsilon_*$ from stellar population models given the
range in galaxy colors.  For early-type galaxies with large bulges, the value of
$\Upsilon_*$ can range from 0.6 to 0.7.  Thus, the systematic variation for
$\Upsilon_*$ is approximately 0.15 for disks and 0.10 for bulges.

Likewise, the gas correction term to convert HI values into total gas mass has two
components of uncertainty, metallicity and the amount of molecular gas in a galaxy.
In the past, we have used a conversion factor ($\eta$) of 1.33 for atomic to total
gas mass that corrects for the abundance of He in low-metallicity systems (high-mass
disks have metallicities near solar, but their gas fractions are small; see
McGaugh 2012).  We have also ignored molecular gas (primarily $H_2$) owing to the
arguments outlined in McGaugh (2012) and observations from McGaugh \& Schombert
(2015).  

In this paper, we will make minor corrections for metallicity and molecular gas using
scaling relations from McGaugh, Lelli \& Schombert (2020).  The gas metallicity causes the gas
correction term, $\eta$, to vary from 1.33 for low-metallicity systems to 1.40 for
galaxies with metallicity near solar (McGaugh 2012).  As the gas-rich systems in our
study are typically low in stellar mass, their metallicities are expected to be low.
In comparison, the high-metallicity galaxies have high $\eta$ values but their gas
component is small.  For this study, we allow $\eta$ to vary with mass/metallicity
following the prescriptions in McGaugh, Lelli \& Schombert (2020); however, this
correction is small (see Table 1).

A similar correction is needed for the contribution of molecular gas, which again is
measured to be very low for low-mass galaxies and higher for high-mass systems.  From
studies of H$_2$ content in disk galaxies, we adopt a scaling relation from total
stellar mass to $M_{H_2}$ (McGaugh, Lelli \& Schombert 2020).  The contribution of
$H_2$ varies from 1\% to 8\% with a mean of 5\% for the SPARC sample.  This is
at the same level as the variation in $\eta$ for the combined gas content.  

To quantify the magnitude of the above variations on the baryon mass, we have listed
six scenarios in Table 1.  {Here we have made a maximum likelihood fit (Lelli \etal
2019) using the BayesLineFit software\footnote{BayesLineFit is available at
http://astroweb.cwru.edu/SPARC/}.}  We fit both the baseline sample of 50 galaxies
with Cepheid and TRGB distances and adequate $V_f$ values (this sample is described
in greater detail in \S2.3) and the SPARC sample (assuming an $H_o = 75$ km s$^{-1}$
Mpc$^{-1}$ for illustration).  We then alter the prescriptions used to calculate the
baryonic mass in the six different ways listed in Table 1 and recalculate the
maximum likelihood fit.  

The largest uncertainty from the stellar population model assumptions arises from
varying $\Upsilon_*$, although the variations in metallicity or molecules for the gas
mass are of the same order of magnitude.  More importantly, neither corrections to
stellar or gas mass have a significant influence on the deduced slope of the bTFR.  All
the slopes are well within the errors of the fitted slope of the redshift-independent
sample of 3.95$\pm$0.16.  For example, a change in the $\eta$ due to metallicity
changes the slope of the main sample by only 0.05.  Large variations in the slope of
the bTFR would make a skewed distribution of residuals from the bTFR, key to deducing an
$H_o$ for each subsample.

In addition, any change in the zeropoint (a shift of the baryon mass to higher or
lower values due to systematic changes in $\Upsilon_*$ or $\eta$) has no effect on
the use of the bTFR as a distance indicator as those shifts are made to the
calibrating galaxies as well as the main sample in identical ways.  Because the
calibrating Cepheid and TRGB sample also covers the same mass range as the full
sample, any minor changes in the slope would also have a negligible effect on the
estimate of $H_o$.

We also list in Table 1 the very small uncertainty due to a zeropoint correction to
the TRGB distance scale (46\% of the calibrating sample) from the discussion in
Freedman \etal (2019).  {A larger systematic shift in the TRGB zero-point would imply
a major offset between Cepheids and TRGB calibrator galaxies in the bTFR plane,
which appears unphysical.}  This zeropoint error is the smallest of all of the
systematic corrections.  The uncertainty in using just TRGB (23 galaxies) or Cepheid
(27 galaxies) calibrators is also small.  As the Cepheid calibrators are primarily
high in mass, this produces a shallower slope to the bTFR ($x=3.70$) compared to the
entire sample.  However, this is still within the errors of all of the other
subsamples and is mitigated by the inclusion of a larger dynamic range for the SPARC
bTFR. 

Errors in $V_f$ are independent of distance, only constrained by the errors in the HI
measurements themselves.  The uncertainty in $V_f$ considers three sources of error:
(1) the random error on each velocity point along the flat part of the rotation
curve, quantifying noncircular motions and kinematic asymmetries between the two
sides of the disk, (2) the dispersion around the estimate of $V_f$ along the rotation
curve, quantifying the actual degree of flatness, and (3) the assumed inclination angle,
which is generally derived from fitting the HI velocity field.  Quantifying the
degree of flatness of the rotation curve and the error of the outer disk inclination
is outlined in Lelli, McGaugh \& Schombert (2016b) and new error calculations are
presented in Lelli \etal (2019) with a mean of 8\% for the sample.  This translates
into a mean error of 0.02 in log $V_f$ on the x-axis of the bTFR.

\subsection{Distances}

Distance is the critical parameter for converting 3.6 apparent magnitudes and HI
fluxes into stellar and gas masses (as both are dependent on $D^2$) and, thereby,
into total baryonic masses ($M_b$).  Within our 125 SPARC galaxies with accurate
RCs, there are 30 galaxies with redshift-independent Cepheid or TRGB distances.  The
Cepheid and TRGB galaxies are listed in Table 2 (TRGB distances and errors are from
Tully \etal 2019, Cepheid distances and errors are from Bhardwaj \etal 2016).  

In addition to the SPARC sample, we have also considered 20 intermediate-mass disk
galaxies from the Ponomareva \etal (2018) study, which also has Cepheid or TRGB
distances (listed in Table 3).  Of the 31 galaxies in the Ponomareva \etal sample
that met our inclination and accurate RC criterion, 11 are already in the SPARC
sample and we have used the SPARC data.  For the remaining 20 galaxies, we have used
their published HI fluxes and $V_f$ values, but have redetermined their stellar mass
values using our own $Spitzer$ 3.6 luminosities, plus our adopted $\Upsilon_*$ values
and gas mass prescriptions.  Their quoted errors are similar to the SPARC sample.
The final sample of 50 galaxies is show in Figure \ref{ctrgb} and is used to
calibrate the bTFR with galaxies having redshift-independent distances (hereafter,
the C/TRGB sample).  {For uniformity, we have only used the distances from Tully \etal
(2019) and Bhardwaj \etal (2016) for both the SPARC and Ponomareva \etal samples.}

To check the internal consistency in the Cepheid versus TRGB methods, there are 41
galaxies in the CosmicFlows-3 database with both Cepheid distances and TRGB
measurements.  Comparison between these Cepheid and TRGB distances indicates an
internal dispersion of 2\% in distance with no obvious systematics (although there
are very few galaxies with C/TRGB distances beyond 10 Mpc).  As discussed in Freedman
\etal (2019), the error in the C/TRGB zeropoint is, at most, 0.05 mag (2\% in
distance), which is consistent with the dispersion between the Cepheid and TRGB methods
from the CosmicFlows-3 dataset.  

For the C/TRGB sample, observational error dominates the error budget compared to
distance error in the $M_b$ values.  The observational error is 0.04 in log $M_b$
whereas a 5\% uncertainty to the C/TRGB distances contributes 0.05 in log $M_b$
($V_f$ being independent of distance).  Thus, the uncertainty in the y-axis of the
bTFR is at most 0.06 in log $M_b$.  And, if the slope is constant from sample to
sample, the limit to our knowledge of the y-intersect is limited by this error and
the sample size.

\subsection{The baseline baryonic Tully-Fisher relation and the flow SPARC sample}

\begin{figure}
\centering
\includegraphics[scale=0.85,angle=0]{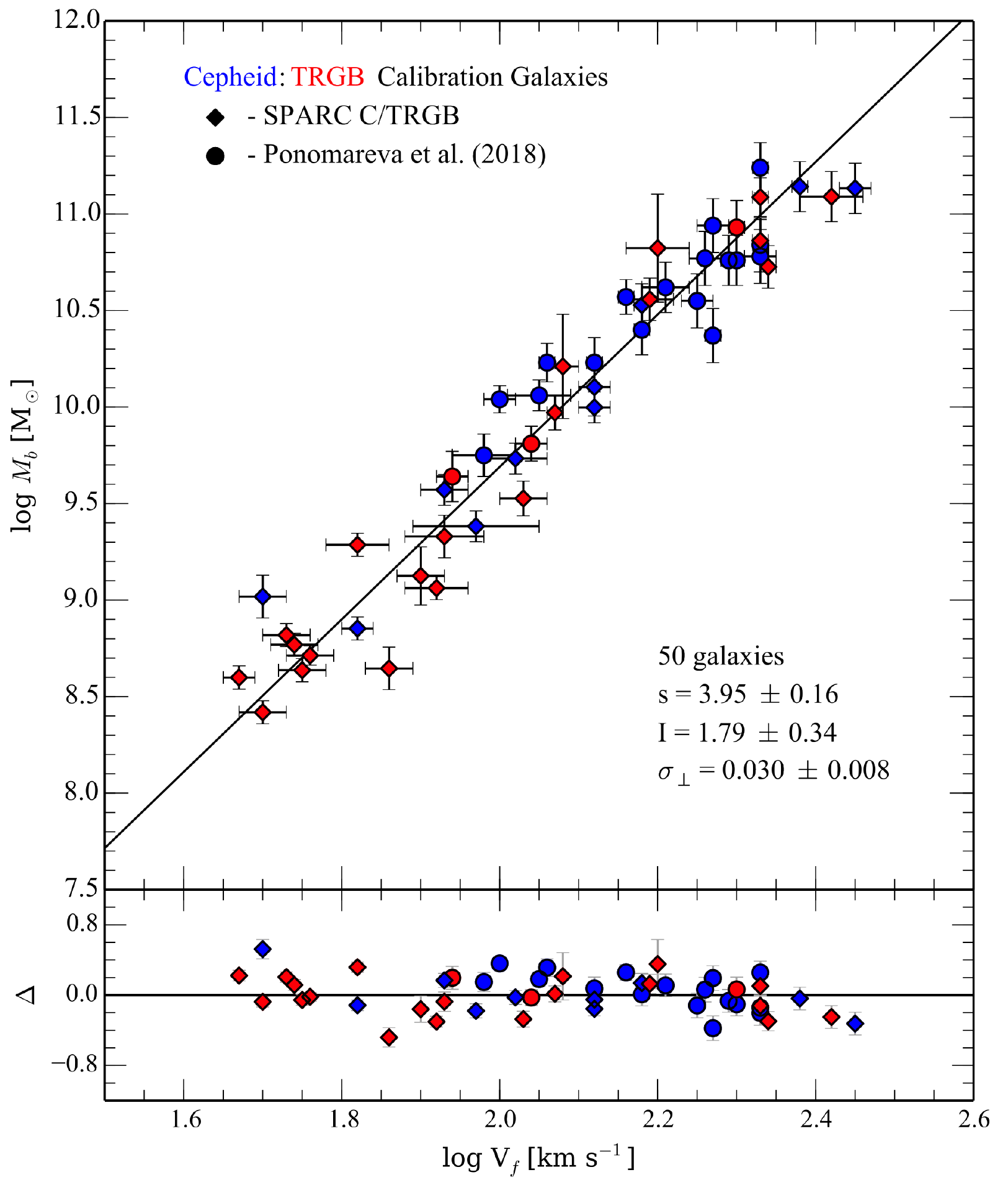}
\caption{\small The bTFR diagram for 30 SPARC galaxies (diamond symbols) and 20
Ponomareva \etal (2018) disks (circular symbols) with C/TRGB distances. TRGB galaxies
are marked in red, Cepheid galaxies in blue.  The baryonic
mass is the sum of the $Spitzer$ 3.6 luminosity converted to stellar mass (see
Schombert, McGaugh \& Lelli 2019) and the total gas mass.  The rotational velocity,
$V_{f}$, is determined directly from HI rotation curves following the techniques outlined in
Lelli \etal (2016).  A maximum likelihood fit is shown and serves as the baseline slope and
zeropoint for comparison to the flow SPARC sample of 95 galaxies.  
}
\label{ctrgb}
\end{figure}

With uncertainties from the previous section in mind, the maximum likelihood fit to
the C/TRGB sample is shown as the solid line in Figure \ref{ctrgb} (also listed in
Table 1).  This slope is consistent, within the errors, to the slope from Lelli \etal
(2019, 3.85 versus 3.95 for the C/TRGB sample).  For a mean error of 0.06 in log
$M_b$ and 0.02 in log $V_f$, the expected scatter for a slope of 3.95 is 0.050 in
log-log space.  The perpendicular residuals have a dispersion of 0.048, exactly what
is expected for uncertainty solely from observational error and intrinsic scatter in
$\Upsilon_*$.  Systematic model uncertainties are well mapped, and they are applied
equally to all of the galaxies in the sample with only slight differences from low to
high-mass due to early-type morphology (i.e., bulge light) and additional molecular
gas for high mass galaxies.  Their effects are more significant on the slope of the
bTFR (as seen in Table 1), but will be irrelevant because our comparison samples will
have the exact same corrections over the same range in galaxy mass.

In our 2016 study, we used a combination of redshift-independent distances, cluster
distances and Hubble flow distances (with an $H_o=73$ km s$^{-1}$ Mpc$^{-1}$) to
calculate the baryonic masses.  Since that study, the CosmicFlows-3 database has been
released (Tully \etal 2019) with an improved flow model (Shaya \etal 2017).
CosmicFlows-3 uses thousands of galaxies and clusters of galaxies to infer flow
deviations from the cosmic expansion velocities owing to peculiar galaxy motions.  The
resulting dynamical model is then reversed with the knowledge of the observed
redshift and position in the sky to deduce the true expansion velocity, which is then
converted into distance using a particular $H_o$.  For our analysis we use a subset
of the inclination-selected SPARC sample of 125 galaxies where we have removed the 30
galaxies used for the C/TRGB calibration, leaving 95 galaxies with good CosmicFlows-3
or Virgo infall velocities (hereafter the flow SPARC sample).  

There is a grouping in the flow SPARC dataset of 26 galaxies at 18 Mpc that
represents the Ursa Major cluster (Verheijen \& Sancisi 2001), where we assumed all
the Ursa Major galaxies to be at the mean cluster distance.  In the interim, the
CosmicFlows-3 project has assigned a mean distance of 17.2 Mpc for Ursa Major using
the luminosity-based TF relation of 35 galaxies.  We adopt this new distance for an $H_o=75$
km s$^{-1}$ Mpc$^{-1}$ with the caveat of adjusting this cluster distance for varying
$H_o$ values.

\begin{figure*}
\centering
\includegraphics[width=\columnwidth,scale=2.0,angle=0]{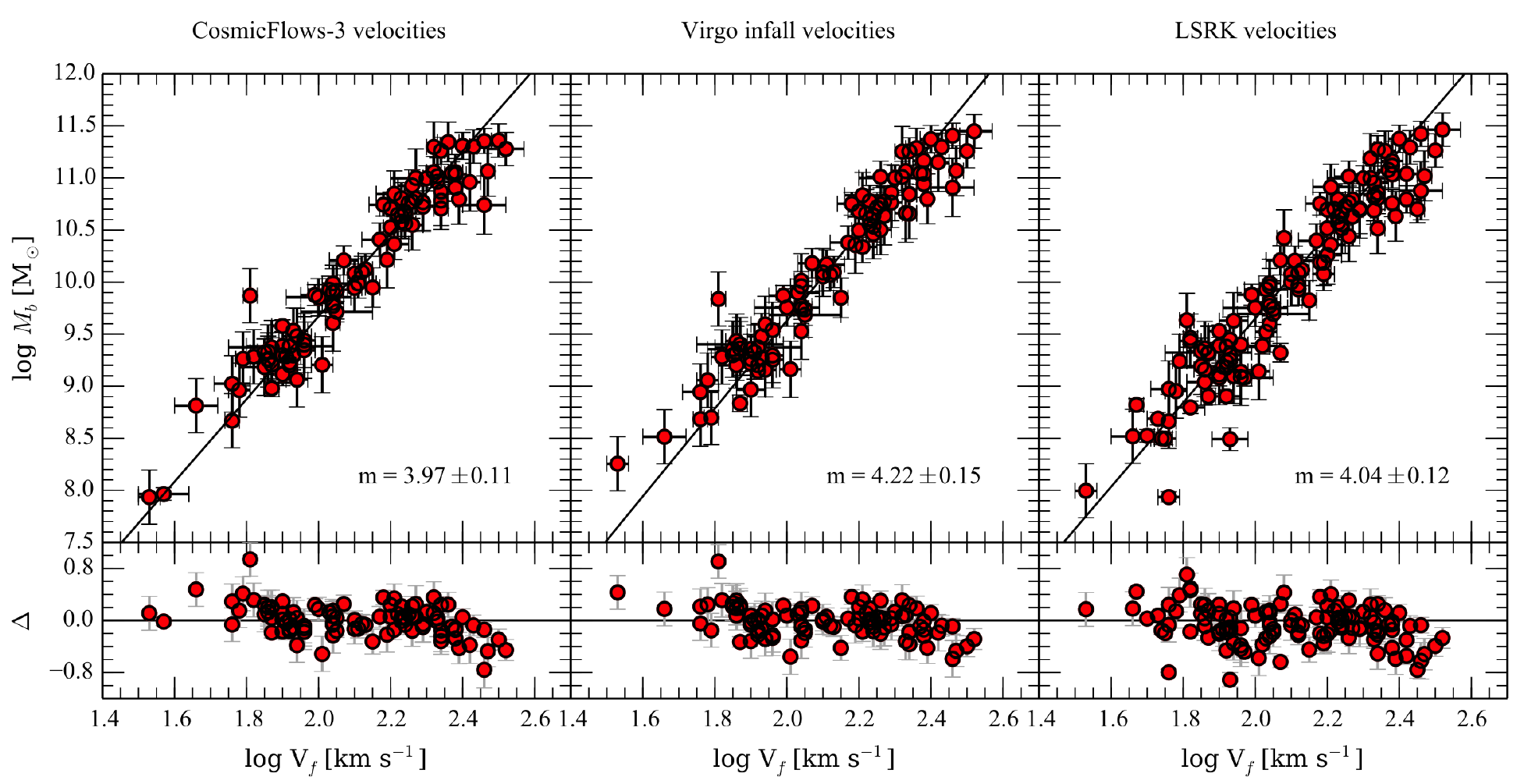}
\caption{\small The bTFR for the flow SPARC galaxies with inclinations
$>$ 30$^\circ$ and flat, outer rotation curves for highly accurate $V_f$ values.
The left panel displays the baryon masses calculated using the CosmicFlows-3
velocities.  The solid line is the maximum likelihood fit.  The
center panel displays the baryon masses calculated for flow velocities based on a
Virgo infall model from Mould
\etal (2000).
There is very little difference in the
bTFR using either CosmicFlows-3 or Mould \etal velocities.  The right panel displays the
resulting bTFR using only local standard of rest (LSRK) velocities. While similar to
the infall distances, the scatter is 15\% larger, from which we conclude that LSRK
velocities are inadequate to deduce $H_o$.  The residuals in $M_b$ from the
C/TRGB fit are shown in the bottom panels.  The formal fits are listed in Table 5.
}
\label{cosmic}
\end{figure*}

The slope of the maximum likelihood fit to the flow SPARC sample (using CosmicFlows-3
velocities; see Figure \ref{cosmic}) is $x$=3.97$\pm$0.12.  The calibrating sample
has an identical slope as the full sample, thus we feel confident in applying the fit
from the C/TRGB galaxies as a baseline comparison for the full sample with varying
$H_o$'s.  In other words, we adopt the same slope of the C/TRGB sample and deduce
$H_o$ from the varying zeropoint on the flow SPARC sample.  

Comparing the CosmicFlows-3 distance to the C/TRGB sample (for an $H_o=75$ km
s$^{-1}$ Mpc$^{-1}$) finds a mean of zero and a standard deviation of 20\% of the
distance.  The Virgo infall models (Mould \etal 2000) find similar means and
dispersions.  As the random errors in the C/TRGB distances are 5\%, this implies most
of the distance uncertainty is in the determination of the flow velocities on the
level of 20\%, in agreement with the uncertainty estimates from Tully \etal (2019).
Given that the uncertainty in the baryon mass of the bTFR diagram is nearly twice
that of the velocity axis, plus the distance errors will only map into the baryon
mass, we then deduce that the dominant term for uncertainty for the flow SPARC sample
is the combined observational errors to $M_b$ plus distance error.  The observational
errors contribute 0.04 on log $M_b$ where distance errors contribute as $D^2$ for
0.16 in log $M_b$.  We assign a combined mean uncertainty of 0.20 in log for the
baryon mass axis.  Again, for a mean error of 0.20 in log $M_b$ and 0.02 in log
$V_f$, the expected scatter for a slope of 3.95 is 0.085 in log-log space. The
perpendicular residuals have a dispersion of 0.067, slightly lower than what is
expected for uncertainty solely from observational error.  Using a slope of 3.97
lowered the dispersion by only a small amount to 0.065.

To explore the robustness of the derived $H_o$ value, we repeat the linear fitting of
the calibrating C/TRGB and the flow SPARC sample using three techniques.  The first,
used in \S2.3, is the maximum likelihood technique considering the orthogonal
intrinsic scatter.  Here the intrinsic scatter is assumed to be Gaussian in the
orthogonal direction to the fitted line.  A second technique is to use the same
maximum likelihood fit but with the intrinsic scatter in the vertical direction.
{Both fitting techniques are implemented in the BayesLineFit
software available at the SPARC website.}  Lastly, we use a least trimmed squares
technique (LTS; Cappellari \etal 2013) using the measured errors in $M_b$ and $V_f$
(the Appendix of Lelli \etal 2019 discusses the mathematical and statistical
properties of these methods and provide basic comparisons).

The results for the three fitting techniques are shown in Table 4.  While the ML
vertical and LTS fits display shallower slopes, the slopes are consistent between the
calibrating C/TRGB and flow SPARC samples (in other words, larger and different
samples produce the same trends between orthogonal, vertical and LTS).  Slightly
different midpoints to the fits will result in a systematic variation in the deduced
$H_o$ values from the orthogonal fit as shown in the table as $\Delta H_o$.  Even
though the uncertainties are greater in the vertical direction, we believe the
orthogonal fits are more representative of the data and the non-negligible error in
$V_f$.  We can assign a systematic error of $\pm$0.5 to the fitting methods.

\section{Results}

The bTFR for the flow SPARC sample (95 galaxies) is shown in Figure \ref{cosmic} using
the CosmicFlows-3 velocities (left panel), the Virgo infall flow model (Mould
\etal 2000, center panel) and only the LSRK velocities
(galactocentric velocity, right panel).  There is very little difference in the
distribution using CosmicFlows-3 versus Virgo infall velocities.  The derived Virgo
infall slope is 4.22, compared to 3.97 for the CosmicFlows-3 values, which is
primarily due to the fact that the flow corrections are small for a majority of the
SPARC sample.  The LSRK galactic center velocities have a similar slope (4.04), but
with 15\% more scatter and the loss of five galaxies with negative redshifts.  

We conclude that distance uncertainties are identical in either method of
correcting for local infall to the observed redshifts and we use the CosmicFlows-3
velocities for our final conclusions.  Using a C/TRGB slope of 3.95, the calibrating
and flow SPARC samples have similar dispersions around that zeropoint.  In other
words, the two samples, with identical sources of error in their kinematics and
photometry, produce the nearly identical slopes to the bTFR whether using redshift
independent C/TRGB distances or CosmicFlows-3 velocities.

A uniform change in the distance model (i.e., $H_o$) will produce a linear shift from
the C/TRGB bTFR (upward for larger distances) such that the bTFR can be used as a
distance scale indicator much like the luminous TF or the Fundamental Plane relation.
In other words, one can reproduce the C/TRGB bTFR zeropoint with the appropriate $H_o$
to convert redshifts into distance to apply to 3.6 luminosities and HI fluxes to
determine the baryon mass.  Comparing the residuals along the $M_b$ axis with the
linear fit to the C/TRGB sample becomes a simple t-test.  In this case, the empirical
correlation of the bTFR is stronger (less scatter) than the Fundamental Plane.  In
addition, errors in the distance apply to each galaxy, rather than a cluster or group
uncertainty as with many traditional applications of the luminous TF relation.  So, increasing
the sample size has a notable effect on the scatter versus adding a new galaxy
cluster to the luminous TF relation.  

A formal match from the C/TRGB bTFR to the flow SPARC sample using CosmicFlows-3
velocities produces a $H_o = 75.1 \pm 2.3$ km s$^{-1}$ Mpc$^{-1}$ using the maximum
likelihood orthogonal fitting method.  This result is relatively independent of bTFR
slope and/or flow model.  Slight changes in the slope (for example, from the 3.9 to
4.1) only widens the dispersion of the residuals from the bTFR, and the differences
between the different distance samples operate only along the $M_b$ axis.  The use of
a slope of 3.8 or 4.1 (the range from the luminous TF relation) has no effect on the mean
normalization and would only increase the error on $H_o$ by 0.5\%.  Likewise, as can
be seen in Figure \ref{cosmic}, using the Virgo infall flow model produces a nearly
identical bTFR fit to the CosmicFlows-3 bTFR (and an $H_o = 74.6$ km s$^{-1}$
Mpc$^{-1}$).

{Of the calibrating sample of 50 galaxies, 23} are TRGB galaxies and any zeropoint
errors would enter through errors in the TRGB method.  The estimated uncertainty on 
$M^{TRGB}_{814}$ is 0.022 (stat) and 0.039 (sys) (primarily through uncertainty to
the distance to the LMC, Riess 2019).  As a numerical experiment, we allowed the
zeropoint for the TRGB galaxies to shift upward and downward by 0.05 mag then refit
the bTFR for the C/TRGB sample.  The results are shown in Table 1.  Due to the fact
that 27 of the 50 galaxies in the C/TRGB sample uses Cepheids as a distance
indicator, the effect on the C/TRGB bTFR is small, only $\pm$0.3 (sys) in the
determination of $H_o$.  In a similar fashion, we examined the effect of different
fitting methods of the C/TRGB sample.  This procedure was discussed in \S2.4 and we
adopt a systematic error in the fitting methods of $\pm$0.5 (sys).

Comparison of the flow models to C/TRGB galaxies finds all of the expected distances
to be within one $\sigma$ of C/TRGB distances with similar dispersions of 20\%.  We
find the CosmicFlows-3 to be the closest match to the C/TRGB distances followed by
the simple Virgo infall then the Virgo plus Great Attractor (Virgo+GA) correction.
The Virgo + Great Attractor + Shapley supercluster produced the same results as the
Virgo+GA model.  As not applying infall corrections raises the scatter by 15\% to 20\%,
we find, unsurprisingly, that LSRK values do not represent the correct value of $H_o$.
In addition, the slope of the bTFR from orthogonal fits using the CosmicFlows-3
velocities is closer to the C/TRGB bTFR slope than using any of the Virgo infall
models.  For this reason, we adopt the CosmicFlows-3 model and the orthogonal fits
for our conclusions.

Using the Virgo infall model consistently finds an $H_o$ value 0.4 km s$^{-1}$
Mpc$^{-1}$ lower than the value found from CosmicFlows-3 velocities across all three
fitting methods.  Using the Virgo+GA and Virgo+GA+Shapley infall models consistently
finds $H_o$ values greater than the baseline by 2.4 km s$^{-1}$ Mpc$^{-1}$.  No
correction for peculiar velocities (LSRK) consistently finds $H_o$ values 2.5 km
s$^{-1}$ Mpc$^{-1}$ lower than the CosmicFlows-3 values.  The standard deviation
between the various flow models is found to be 1.5 km s$^{-1}$ Mpc$^{-1}$, which we
adopt as the systematic variation for our flow model.  As this value dominates
the other systematics in the fitting methods or zeropoint errors, we use this value
to indicate the expected systematic uncertainty in our measurements.  Thus, we adopt a
formal fit to the flow SPARC dataset of $H_o = 75.1 \pm 2.3$ (stat) $\pm 1.5$ (sys)
km s$^{-1}$ Mpc$^{-1}$.

\begin{figure*}
\centering
\includegraphics[width=\columnwidth,scale=4.0,angle=0]{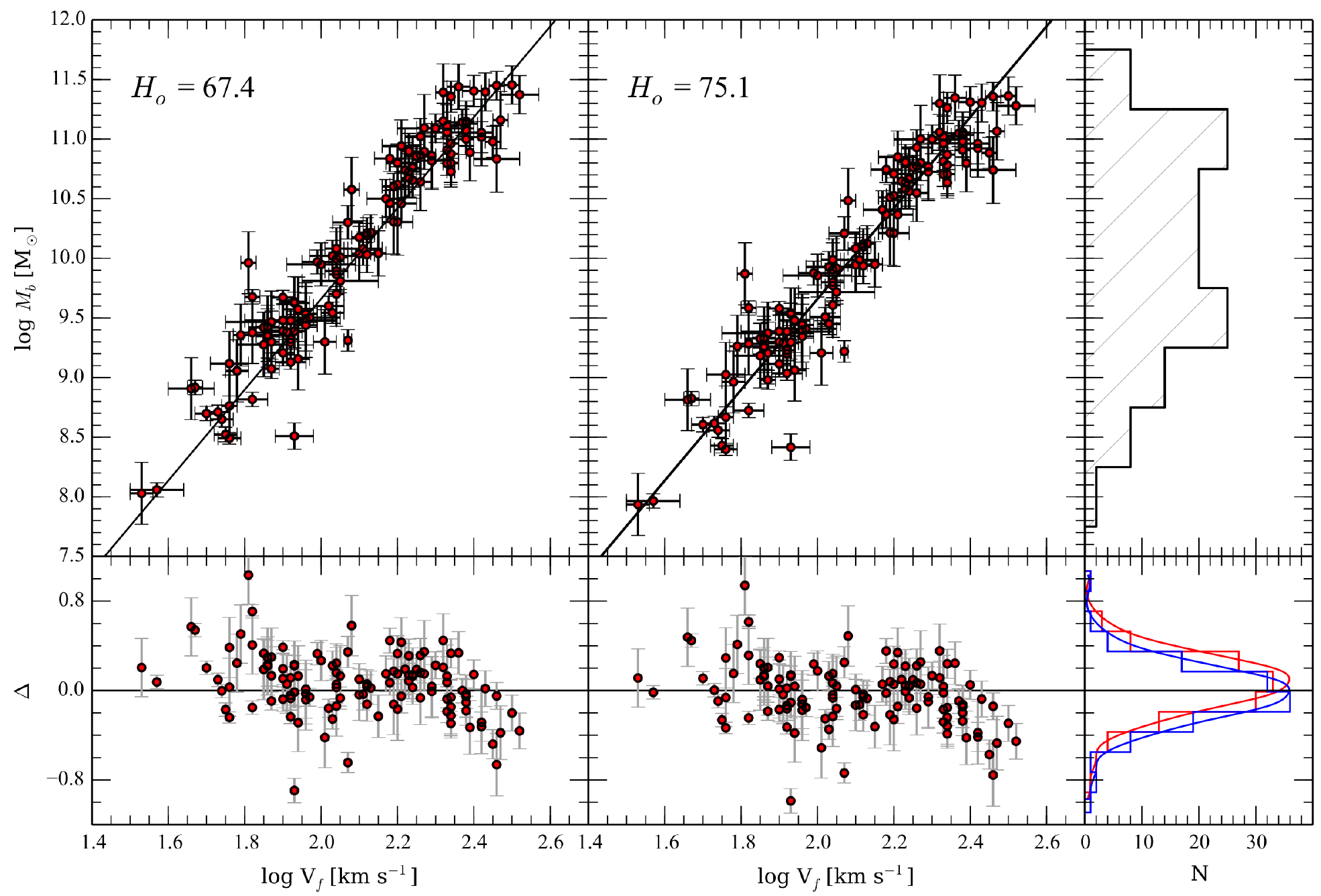}
\caption{\small The bTFR for the flow SPARC galaxies, using CosmicFlows-3 velocities,
with a $H_o=67.4$ km s$^{-1}$ Mpc$^{-1}$ (left panel) and $H_o=75.1$ km s$^{-1}$
Mpc$^{-1}$ (right panel).  The solid line is the baseline fit from Figure
\ref{ctrgb}.  Residuals are plotted in the bottom panels.  Histograms of the range in
$M_b$ and residuals are shown on the right (blue for 75.1, red for 67.4).  As $V_f$ is independent of
distance, changing $H_o$ has the effect of increasing the baryon mass by $\Delta
H_o^2$.  $H_o$ values less than 70.5 are ruled out at the 95\% confidence level.
}
\label{shift}
\end{figure*}

To test the significance level of the orthogonal fit to competing values of $H_o$, we
recalculate the flow SPARC sample baryonic masses using distances given by an
$H_o=67.4$ km s$^{-1}$ Mpc$^{-1}$.  This is the value of $H_o$ found by the Planck
mission (Planck Collaboration \etal 2018) fitting CMB anisotropies with the base
$\Lambda$CDM cosmology.  The resulting bTFRs are shown in Figure \ref{shift}
(residuals in the bottom panel).  Assuming the linear fit to the calibrating C/TRGB
sample (shown as a solid line in Figure \ref{shift}), we can perform a simple t-test
analysis on the distribution of perpendiculars from the linear fit.  For the C/TRGB
$H_o=75.1$ km s$^{-1}$ Mpc$^{-1}$ and $H_o=67.4$ km s$^{-1}$ Mpc$^{-1}$ samples we
have, respectively, mean perpendiculars of 0.0 (by definition), $-$0.001 and $-$0.024 with standard
deviations of 0.057 for the calibrating sample and 0.069 for the $H_o$ samples.  The
C/TRGB and $H_o=75.1$ km s$^{-1}$ Mpc$^{-1}$ samples are in agreement at a high
level, the $H_o=67.4$ km s$^{-1}$ Mpc$^{-1}$ sample is rejected at the 99.98\% level.
In fact, all values of $H_o$ below 70.5 are rejected at the 95\% level.

\section{Conclusions}

As outlined in the Introduction, the baryonic Tully-Fisher relation is one of the
strongest {\it empirical} correlations in extragalactic astronomy.  While there exist
many other correlations between galaxy characteristics with similar scatter (such as
scaling relations between effective radius and surface brightness in ellipticals; see
Schombert 2017), other galaxy correlations often involve coupled parameters measured
by a similar procedure (e.g., photometry).  The bTFR involves two distinct parameters,
rotation velocity and baryonic mass, both measured by independent methods (HI
observations provide both the HI mass and the rotation velocities, but the former
comes from the total intensity map and the latter from the velocity map, so they are
effectively independent observables).  With the advent of accurate rotation curves,
the determination of $V_f$ has very little uncertainty and is mapped to the very edge
of the baryonic extent of a galaxy.  In addition, the behavior of $V_f$ with respect
to past measures of rotation velocity is now well known (see Lelli \etal 2019;
Ponomareva \etal 2018). 

Likewise, the use of near-IR luminosities to determine the stellar mass component of
the baryonic mass has minimized the largest source of uncertainty on that axis of the
bTFR, the value of $\Upsilon_*$ {(see Zaritsky \etal 2014)}.  The error budget in
$M_b$, as applied to as a new distance indicator, is now dominated by the observational
error in the HI fluxes and 3.6 photometry, which approaches 10\% each.  This means the
baryonic mass, which is obtained without reference to kinematics, is determined to a
higher degree of accuracy than could be obtained simply from the kinematics.  The error
on this axis is completely dominated by uncertainty in distance, which in turn makes
it an ideal distance scale indicator when calibrated with redshift-independent
samples.  Our analysis herein results in a high value of $H_o$ near 75 km s$^{-1}$
Mpc$^{-1}$.

It is becoming increasingly obvious that values of $H_o$, determined from empirical
astronomical correlations differ significantly from values determined from
expectations estimated by cosmological model fits (e.g., fits to the angular power
spectrum of the CMB fluctuations).  There is a long history in observational
astronomy of constructing a distance scale to determine the local $H_o$ independent of
any underlying astrophysics.  In fact, the best techniques rely on as little modeling
as possible.  Where model values are used (for example, $\Upsilon_*$ in determining
stellar mass from 3.6 luminosities), those values are highly constrained by limits
based on knowledge of galaxy colors and past SFRs.

However, one cannot ignore that the model fits to the CMB under a $\Lambda$CDM
cosmology have been extremely successful at explaining the details of these
observations of the early universe.  For example, the concordance $\Lambda$CDM model
is in agreement with measurements of anisotropies in the temperature and polarization
of the CMB (Planck Collaboration \etal 2018) as well as fluctuations in the baryonic
acoustic oscillations (Alam \etal 2017).  The temperature and polarization anisotropy
spectra are well fit by a six-parameter model that does not give a direct measurement
of $H_o$, but provide an inferred value of $67.4\pm0.5$ km s$^{-1}$ Mpc$^{-1}$.  

An important distinction is that observations of the CMB do not measure $H_o$
directly, but rather predict what the value of $H_o$ should be given a specific model
of the expanding universe with cosmological parameters in a $\Lambda$CDM framework.
This framework also makes predictions on the power spectrum, polarization and
anisotropies (also expressed through various cosmological parameters).  The actual
measured values of $H_o$, through distance ladder techniques, do not agree with this
deduced value.

In this study, we present another empirical method to deduce $H_o$.  We calibrate the
bTFR using 50 galaxies with Cepheid and/or TRGB distances and apply this relation to
another 95 galaxies from the SPARC sample to deduce the value of $H_o$.  Due to the
nature of the observations, only the baryonic mass axis is sensitive to distance.
Leveraging the fit to the redshift independent calibrators, we find that $H_o = 75.1$
km s$^{-1}$ Mpc$^{-1}$ ruling out all values below 70 with a 95\% degree of
confidence.

High values of $H_o$ near 75 km s$^{-1}$ Mpc$^{-1}$ continue to be more appropriate
when involving empirical, observational issues of galaxy distance, while lower values
of $H_o$ are required for the framework that involves the CMB and the physics of the
early universe.  These are, in a real sense, separate chains of deductive reasoning.
And, while in a rule-driven universe, $H_o$ should be connected from the early
universe to today, there exist many explanations that do not require that to be true
(Verde, Treu \& Riess 2019).  In addition, there is increasing reason to doubt the
CDM paradigm on galaxy scales (see McGaugh, Lelli \& Schombert 2016; Bullock \&
Boylan-Kolchin 2017) despite its success on cosmological scales

The future use of the bTFR as a distance indicator is encouraging due to the fact
that the intrinsic bTFR scatter in the C/TRGB sample and the flow SPARC sample are similar,
despite the difference in sample size.  This indicates that additional calibrating
galaxies will significantly improve the slope and zeropoint to the bTFR while
additional flow galaxies will severely push down both random and systematic
errors.  {In fact, the current dominating systematic uncertainty ($\pm$1.5 km s$^{-1}$
Mpc$^{-1}$) depends on how the 95 flow SPARC galaxies sample specific regions of the
nearby universe, where flow velocities can differ by more than the mean global
difference from different flow models (e.g.  CosmicFlows-3 versus Virgo Infall).
Additional flow galaxies will allow us to sample a larger volume of the nearby
universe, where different flow models have, on average, smaller differences.}

\section*{Acknowledgements}
We thank Brent Tully and an anonymous referee for comments to improve the text.
Software for this project was developed under NASA's AIRS and ADAP Programs. This work
is based in part on observations made with the Spitzer Space Telescope, which is
operated by the Jet Propulsion Laboratory, California Institute of Technology under a
contract with NASA.  Support for this work was provided by NASA through an award
issued by JPL/Caltech. Other aspects of this work were supported in part by NASA ADAP
grant NNX11AF89G and NSF grant AST 0908370. As usual, this research has made use of
the NASA/IPAC Extragalactic Database (NED) which is operated by the Jet Propulsion
Laboratory, California Institute of Technology, under contract with the National
Aeronautics and Space Administration.

\begin{deluxetable}{lcccc}
\tablecolumns{5}
\small
\tablewidth{0pt}
\tablecaption{Systematic Error Budget}

\tablehead{
\\
\colhead{case} & \colhead{log $A$} & \colhead{$x$} & \colhead{$\Delta{H_o}$} & \colhead{Notes} \\
}
\startdata
baseline            & 1.79$\pm$0.34 & 3.95$\pm$0.16 &  --  & 50 C/TRGB galaxies \\
Low $\Upsilon_*$    & 1.86$\pm$0.24 & 3.90$\pm$0.11 & $-$0.4 & $\Upsilon_*$=0.4 \\
High $\Upsilon_*$   & 1.63$\pm$0.24 & 4.02$\pm$0.11 & +0.2 & $\Upsilon_*$=0.6 \\
No Molecules        & 1.80$\pm$0.24 & 3.93$\pm$0.11 & $-$0.3 & \\
No Heavy Elements   & 1.74$\pm$0.24 & 3.96$\pm$0.11 & +0.1 & $\eta$=1.33 \\
High TRGB zeropoint & 1.81$\pm$0.33 & 3.94$\pm$0.16 & +0.3 & +0.05 mag \\
Low TRGB zeropoint  & 1.77$\pm$0.34 & 3.96$\pm$0.16 & $-$0.3 & $-$0.05 mag \\

\enddata
\tablecomments{The bTFR is in the form of $M_b = A{V_f}^x$.}

\end{deluxetable}
\clearpage

\begin{deluxetable}{lr@{.}lcr@{.}ll}
\tablecolumns{5}
\small
\tablewidth{0pt}
\tablecaption{SPARC Cepheids/TRGB Calibrating Galaxies}

\tablehead{
\\
\colhead{Galaxy} & \multicolumn{2}{c}{D} & \colhead{log $V_f$} & \multicolumn{2}{c}{log $M_b$} & \colhead{Distance} \\
 & \multicolumn{2}{c}{(Mpc)} & \colhead{(km s$^{-1}$)} & \multicolumn{2}{c}{($M_{\odot}$)} & \colhead{Method} \\
}
\startdata
D631-7 & 7&87$\pm$0.20 & 1.76$\pm$0.03 & 8&71$\pm$0.05 & TRGB \\
DDO154 & 4&04$\pm$0.15 & 1.67$\pm$0.02 & 8&60$\pm$0.06 & TRGB \\
DDO168 & 4&25$\pm$0.20 & 1.73$\pm$0.03 & 8&82$\pm$0.06 & TRGB \\
IC2574 & 3&89$\pm$0.14 & 1.82$\pm$0.04 & 9&29$\pm$0.06 & TRGB \\
NGC0024 & 7&67$\pm$0.32 & 2.03$\pm$0.03 & 9&53$\pm$0.09 & TRGB \\
NGC0055 & 1&89$\pm$0.05 & 1.93$\pm$0.03 & 9&57$\pm$0.08 & Cepheids \\
NGC0247 & 3&41$\pm$0.14 & 2.02$\pm$0.04 & 9&73$\pm$0.08 & Cepheids \\
NGC0300 & 1&91$\pm$0.06 & 1.97$\pm$0.08 & 9&38$\pm$0.08 & Cepheids \\
NGC0891 & 9&12$\pm$0.34 & 2.33$\pm$0.01 & 10&86$\pm$0.11 & TRGB \\
NGC1705 & 5&51$\pm$0.20 & 1.86$\pm$0.03 & 8&65$\pm$0.11 & TRGB \\
NGC2366 & 3&34$\pm$0.09 & 1.70$\pm$0.03 & 9&02$\pm$0.11 & Cepheids \\
NGC2403 & 3&18$\pm$0.09 & 2.12$\pm$0.02 & 9&99$\pm$0.08 & Cepheids \\
NGC2683 & 8&59$\pm$0.36 & 2.19$\pm$0.03 & 10&56$\pm$0.11 & TRGB \\
NGC2841 & 14&60$\pm$0.47 & 2.45$\pm$0.02 & 11&13$\pm$0.13 & Cepheids \\
NGC2915 & 4&29$\pm$0.20 & 1.92$\pm$0.04 & 9&06$\pm$0.06 & TRGB \\
NGC2976 & 3&63$\pm$0.13 & 1.93$\pm$0.05 & 9&33$\pm$0.11 & TRGB \\
NGC3109 & 1&30$\pm$0.03 & 1.82$\pm$0.02 & 8&85$\pm$0.06 & Cepheids \\
NGC3198 & 13&40$\pm$0.55 & 2.18$\pm$0.01 & 10&53$\pm$0.11 & Cepheids \\
NGC3741 & 3&23$\pm$0.12 & 1.70$\pm$0.03 & 8&42$\pm$0.06 & TRGB \\
NGC3972 & 20&80$\pm$0.67 & 2.12$\pm$0.02 & 10&10$\pm$0.15 & Cepheids \\
NGC4214 & 2&93$\pm$0.11 & 1.90$\pm$0.03 & 9&12$\pm$0.15 & TRGB \\
NGC4559 & 8&43$\pm$0.66 & 2.08$\pm$0.02 & 10&21$\pm$0.27 & TRGB \\
NGC5005 & 18&37$\pm$1.27 & 2.42$\pm$0.04 & 11&09$\pm$0.13 & TRGB \\
NGC5907 & 17&10$\pm$0.71 & 2.33$\pm$0.01 & 11&09$\pm$0.10 & TRGB \\
NGC6503 & 6&25$\pm$0.29 & 2.07$\pm$0.01 & 9&97$\pm$0.09 & TRGB \\
NGC6946 & 6&72$\pm$0.50 & 2.20$\pm$0.04 & 10&82$\pm$0.28 & TRGB \\
NGC7331 & 13&90$\pm$0.51 & 2.38$\pm$0.01 & 11&14$\pm$0.13 & Cepheids \\
UGC01281 & 5&27$\pm$0.24 & 1.74$\pm$0.03 & 8&77$\pm$0.06 & TRGB \\
UGCA442 & 4&37$\pm$0.20 & 1.75$\pm$0.03 & 8&64$\pm$0.06 & TRGB \\
\enddata

\end{deluxetable}
\clearpage

\begin{deluxetable}{lr@{.}lcr@{.}ll}
\tablecolumns{5}
\small
\tablewidth{0pt}
\tablecaption{Ponomareva \etal Cepheids/TRGB Calibrating Galaxies}

\tablehead{
\\
\colhead{Galaxy} & \multicolumn{2}{c}{D} & \colhead{log $V_f$} & \multicolumn{2}{c}{log $M_b$} & \colhead{Distance} \\
 & \multicolumn{2}{c}{(Mpc)} & \colhead{(km s$^{-1}$)} & \multicolumn{2}{c}{($M_{\odot}$)} & \colhead{Method} \\
}
\startdata
NGC0253 & 3&56$\pm$0.13 & 2.30$\pm$0.01 & 10&72$\pm$0.14 & TRGB \\
NGC0925 & 8&91$\pm$0.28 & 2.06$\pm$0.01 & 10&14$\pm$0.10 & Cepheids \\
NGC1365 & 17&70$\pm$0.81 & 2.33$\pm$0.01 & 11&04$\pm$0.13 & Cepheids \\
NGC2541 & 11&50$\pm$0.47 & 2.00$\pm$0.02 & 9&89$\pm$0.07 & Cepheids \\
NGC3031 & 3&61$\pm$0.09 & 2.33$\pm$0.02 & 10&71$\pm$0.14 & Cepheids \\
NGC3319 & 13&00$\pm$0.53 & 2.05$\pm$0.04 & 9&91$\pm$0.08 & Cepheids \\
NGC3351 & 10&40$\pm$0.28 & 2.25$\pm$0.02 & 10&47$\pm$0.14 & Cepheids \\
NGC3370 & 26&10$\pm$0.72 & 2.18$\pm$0.01 & 10&27$\pm$0.13 & Cepheids \\
NGC3621 & 6&72$\pm$0.18 & 2.16$\pm$0.01 & 10&38$\pm$0.09 & Cepheids \\
NGC3627 & 9&03$\pm$0.29 & 2.26$\pm$0.02 & 10&68$\pm$0.14 & Cepheids \\
NGC4244 & 4&61$\pm$0.19 & 2.04$\pm$0.02 & 9&68$\pm$0.09 & TRGB \\
NGC4258 & 7&31$\pm$0.16 & 2.30$\pm$0.01 & 10&72$\pm$0.13 & Cepheids \\
NGC4414 & 17&80$\pm$0.74 & 2.27$\pm$0.02 & 10&80$\pm$0.14 & Cepheids \\
NGC4535 & 16&10$\pm$0.66 & 2.29$\pm$0.01 & 10&69$\pm$0.13 & Cepheids \\
NGC4536 & 14&60$\pm$0.60 & 2.21$\pm$0.03 & 10&47$\pm$0.13 & Cepheids \\
NGC4605 & 5&54$\pm$0.25 & 1.94$\pm$0.02 & 9&52$\pm$0.13 & TRGB \\
NGC4639 & 22&00$\pm$0.71 & 2.27$\pm$0.01 & 10&27$\pm$0.14 & Cepheids \\
NGC4725 & 12&50$\pm$0.46 & 2.33$\pm$0.01 & 10&73$\pm$0.14 & Cepheids \\
NGC5584 & 22&40$\pm$0.72 & 2.12$\pm$0.01 & 10&11$\pm$0.13 & Cepheids \\
NGC7814 & 14&39$\pm$0.50 & 2.34$\pm$0.01 & 10&73$\pm$0.11 & TRGB \\
NGC7793 & 3&58$\pm$0.11 & 1.98$\pm$0.04 & 9&62$\pm$0.11 & Cepheids \\
\enddata

\end{deluxetable}

\begin{deluxetable}{cccl}
\tablecolumns{4}
\small
\tablewidth{0pt}
\tablecaption{C/TRGB Fitting Results}

\tablehead{
\\
\colhead{log $A$} & \colhead{$x$} & \colhead{$\Delta{H_o}$} & \colhead{Fit} \\
}
\startdata
1.78$\pm$0.35 & 3.95$\pm$0.17 & --   & ML orthogonal \\
2.29$\pm$0.31 & 3.71$\pm$0.15 & $-$0.3 & ML vertical \\
2.34$\pm$0.03 & 3.69$\pm$0.12 & $-$1.0 & LTS \\
\enddata

\end{deluxetable}

\begin{deluxetable}{lcl}
\tablecolumns{3}
\small
\tablewidth{0pt}
\tablecaption{Flow SPARC Fits}

\tablehead{
\\
\colhead{$x$} & \colhead{$H_o$} & \colhead{flow model} \\
}
\startdata
3.97$\pm$0.11 & 75.1$\pm$2.3 & CosmicFlows-3 \\
4.22$\pm$0.15 & 74.6$\pm$2.4 & Virgo Infall \\
4.01$\pm$0.11 & 77.5$\pm$2.4 & Virgo+GA Infall \\
4.04$\pm$0.12 & 72.8$\pm$2.7 & LSRK \\
\enddata

\end{deluxetable}

\end{document}